# The Band Excitation Method in Scanning Probe Microscopy for Rapid Mapping of Energy Dissipation on the Nanoscale

Stephen Jesse,[1] Sergei V. Kalinin,[1,2,*] Roger Proksch,[3] A.P. Baddorf,[1,2] and B.J. Rodriguez[1,2]

[1] Materials Science and Technology Division, Oak Ridge National Laboratory,
Oak Ridge, TN 37831

[2] Center for Nanophase Materials Sciences, Oak Ridge National Laboratory,
Oak Ridge, TN 37831

[3] Asylum Research, 6310 Hollister Ave., Santa Barbara, CA 93117

## Abstract

Mapping energy transformation pathways and dissipation on the nanoscale and understanding the role of local structure on dissipative behavior is a grand challenge for imaging in areas ranging from electronics and information technologies to efficient energy production. Here we develop a novel Scanning Probe Microscopy (SPM) technique in which the cantilever is excited and the response is recorded over a band of frequencies simultaneously rather than at a single frequency as in conventional SPMs. This band excitation (BE) SPM allows very rapid acquisition of the full frequency response at each point

[*] Corresponding author, sergei2@ornl.gov

(i.e. transfer function) in an image and in particular enables the direct measurement of energy dissipation through the determination of the $Q$-factor of the cantilever-sample system. The BE method is demonstrated for force-distance and voltage spectroscopies and for magnetic dissipation imaging with sensitivity close to the thermomechanical limit. The applicability of BE for various SPMs is analyzed, and the method is expected to be universally applicable to all ambient and liquid SPMs.

# I. Introduction

Energy transformations and the inevitable dissipation associated with it are integral components of the physical world. Development of the science of energy dissipation at its fundamental length scales will have enormous implications on such varied technologies as energy production and utilization[1] and in nanoscale electronic applications and information technologies.[2] Often, macroscopic dissipation has its origin in disperse, highly localized, low-dimensional centers. For instance, transport properties in metals and semiconductors are controlled by scattering at impurities, energy losses in magnetization reversal processes are determined by magnetoacoustic phonon generation associated with domain wall motion and depinning, and energy losses during viscoelastic processes are related to crystal defect motion. Understanding of atomistic dissipation mechanisms and improved engineering of materials and device strategies to minimize energy losses necessitate the development of techniques capable of imaging and characterizing nanoscale dissipative processes on the level of a single dislocation, structural defect, or dopant atom.

Dissipation in materials and devices on the macroscopic scale is easily accessible through direct measurements. The area of the hysteresis loop in ferroelectric or magnetic measurements provides a measure of irreversible work in the system. Similarly, dissipated power can be determined from current-voltage measurements for electric dissipation and loss modulus measurements for mechanical dissipation. Finally, heat generation in a system can be measured to provide information on dissipated energy. However, these macroscale measurements of collective phenomena are not easily extendable to the nanoscale.

Here, we introduce a novel excitation and measurement mode (band excitation, or BE) that allows rapid mapping of energy dissipation on the nanoscale. BE utilizes a non-sinusoidal excitation signal having a finite amplitude over a selected band in frequency space that substitutes the sinusoidal excitation in standard Scanning Probe Microscopies. The principles of energy dissipation measurement and the limitations of classical SPM detection modes are discussed in Section II. The principles of band excitation and experimental implementation of BE are summarized in Sections III and IV respectively. The BE force-distance and voltage spectroscopies are presented in Section V, and BE Magnetic Dissipation Force Microscopy is illustrated in Section VI. The applicability and limitations of the BE method for existing SPM modes are discussed in Section VII.

## II. Energy Dissipation Measurements in SPM

Scanning probe microscopy (SPM), well established for the measurement of topography and forces on the nanoscale, provides a potential strategy for local dissipation measurement.[3,4,5] In this, the SPM tip concentrates the probing field to the nanometer level, and the cantilever acts as an energy dissipation sensor. The energy dissipated due to tip-surface interactions is determined using power balance as $P_{diss} = P_{drive} - P_0$, where $P_{drive}$ is the power provided to the probe by an external driving source, and $P_0$ is the sum of intrinsic losses due to cantilever damping by the surroundings and within the cantilever material. The external power can be determined from the cantilever dynamics as $P_{drive} = \langle F\dot{z} \rangle$, where $F$ is the force acting on the probe, $\dot{z}$ is the experimentally measured probe velocity, with the average taken along the probe tip trajectory. The intrinsic losses within the material and due to the hydrodynamic damping by ambient, $P_0$, are determined by calibration at a reference position, $P_{diss} = 0$.

The dynamic behavior of the cantilever weakly interacting with the surface in the vicinity of the resonance can be well approximated by a simple harmonic oscillator (SHO) model described by three independent parameters, namely resonant frequency, $\omega_0$, amplitude at the resonance, $A_0$, and quality factor, $Q$, as

$$A(\omega) = \frac{A^{max}\omega_o^2}{\sqrt{(\omega^2 - \omega_o^2)^2 + (\omega\omega_o/Q)^2}} \quad \text{and} \quad \tan(\varphi(\omega)) = \frac{\omega\omega_o/Q}{\omega^2 - \omega_o^2} \quad (1a,b)$$

From these, $\omega_0$ is related to the tip-surface force gradient, $A_0$ to the driving force, and $Q$ to the energy dissipation.[6]

For constant frequency operation, seminal work by Cleveland *et al.*[7] and Garcia[8] has related energy loss to the phase shift of a vibrating cantilever. Dissipative tip-surface interactions can be probed via measurement of the amplitude, $A$, and phase, $\varphi$, of the cantilever driven mechanically with amplitude, $A_d$, at a constant frequency, $\omega$, as

$$P_{tip} = \frac{1}{2}\frac{kA^2\omega}{Q_0}\left[\frac{Q_0 A_d \sin\varphi}{A} - \frac{\omega}{\omega_0}\right] \quad (2)$$

where $\omega_0$ is the resonance frequency of the cantilever with spring constant, $k$, and the quality factor in free space, $Q_0$.

The emergence of frequency tracking techniques[9] provides another means to determine dissipation. In this, the cantilever is driven at constant amplitude near the resonance frequency, the response amplitude is measured, and by assuming that changes in the signal strength are proportional to the *Q*-factor, dissipation in the system can be ascertained. In this case,

$$P_{tip} = \frac{1}{2}kA^2\omega_0\left[\frac{1}{Q_0} - \frac{1}{Q}\right] \quad (3)$$

and $Q$ is the quality factor in the vicinity of the surface. Experimentally, $Q$ is determined using an additional feedback loop that maintains the oscillation amplitude constant by adjusting the driving amplitude, $Q = A/A_d$. These approaches were implemented by several groups to study magnetic dissipation,[10,11] electrical dissipation,[12,13] and mechanical dissipation on atomic[14,15] and molecular levels.[16]

Notably, in a standard single-frequency SPM experiment the number of independent parameters defining the cantilever dynamics (i.e. 3 SHO parameters) exceeds the number of experimentally observed variables (e.g., amplitude and phase), precluding direct measurement of dissipation. For acoustically driven systems, the constant driving force, $F = const$, provides an additional constraint required to determine 3 independent SHO parameters from two experimentally accessible quantities [Fig. 1 (a)]. However, Eqs. (2, 3) are no longer valid for techniques where the driving signal is position, time, or frequency dependent, $F \neq const$. For example, in Kelvin probe force microscopy (KPFM), the driving force, i.e., the capacitive tip-surface interaction, is proportional to both the local work function and the periodic voltage applied to the tip. Hence, variations in the signal strength are due both to work function variations and dissipation, and these effects cannot be separated unambiguously [Fig. 1 (b)]. Similarly, in atomic force acoustic microscopy and piezoresponse force microscopy, which are used to address local mechanical and electromechanical properties, variations in the local response cannot be unambiguously distinguished from dissipation.

Even in techniques utilizing constant excitation signals, non-linearities in the tip-surface interaction result in the creation of higher-harmonics which can cause confusion between information about dissipation and other properties.[17] Furtermore, dissipation measurements are extremely sensitive to SPM electronics. For example, small deviations in the phase set-point from the resonance condition in frequency tracking techniques result in major errors in the measured dissipation energy. Most importantly, implementation of these techniques requires the calibration of the frequency response of the piezoactuator driving the cantilever.[18] In the absence of such calibration, the images often demonstrate abnormal cantilever-dependent contrast. All together, these factors contribute to a relative paucity of studies on dissipation processes in SPM.

This limited applicability of SPM to dissipation measurements is a direct consequence of the fact that traditional SPM excites and samples the response at a single frequency at a time. This allows fast imaging and high signal levels, but information about the frequency-dependent response, and hence dissipation and energy transfer, is not probed. At the other extreme, spectroscopic techniques excite and sample over all Fourier space (up to the bandwidth limit of the electronics), but the response amplitude is necessarily small since the excitation energy is spread over all frequencies.[19] Finally, response in the vicinity of the resonance can be probed using frequency sweeps. However in this case, homodyne detection implemented in standard lock-in techniques results in significant phase and amplitude errors and information loss if the relaxation time of the oscillator exceeds the residence time at each frequency. This necessitates long acquisition times to achieve adequate signal to noise ratios, incompatible with 1-30 ms/pixel data acquisition times required for practical SPM imaging.

### III. Principles of Band Excitation Method

Here, we develop and implement a method based on an adaptive, digitally synthesized signal that excites and detects within a band of frequencies over a selected frequency range simultaneously.[20] This approach takes advantage of the fact that only selected regions of Fourier space contain information of any practical interest, for instance in the vicinity of resonances. Instead of a simple sinusoidal excitation, the BE method developed here uses a signal having a predefined amplitude and phase content. The generic process is illustrated in Fig. 2. The signal is generated to have the predefined Fourier amplitude density in the frequency band of interest and inverse Fourier transformed to generate excitation signal in time domain. Resulting complex waveform is used to excite the cantilever electrically, acoustically, or magnetically. The cantilever response to the BE drive is measured and Fourier transformed to yield the amplitude- and phase-frequency curves and is stored at each point in the image as a 3D [ $A(\omega)$ and $\theta(\omega)$ at each point] data sets. The ratio of the response and excitation signals yields the transfer function of the system.

The applicability of BE is analyzed as following. The point spacing in the frequency domain is $\Delta f = 1/T$, where $T$ is the pulse duration (equal to pixel time, ~20 ms for 0.4 Hz scan rate at 128 points/pixel). For a resonance frequency of $\omega_0$ = 150 kHz and $Q$-factor of ~ 200, the width of the resonance peak is ~750 Hz, allowing for sufficient sampling of the peak in the frequency domain (15 points above the half-max). The sampling efficiency increases for lower $Q$-factors (e.g., imaging in liquids) and higher resonance frequencies (contact modes and stiff cantilevers). Remarkably, the parallel detection of the BE method implies that the total number of frequency points (i.e., the width of the band in the Fourier space) can be arbitrary large, with the cost being the signal/noise ratio. Typically for a single peak tracking, the frequency band is chosen such that the intensity factor, defined as $I_{det} = \int A(\omega) d\omega / A_{max} \Delta \omega$, where the integral is taken over the frequency band of width $\Delta \omega$, is $I_{det}$ ~ 0.2-0.7. Alternatively, the excitation signal can be tailored to provide a higher excitation level away from the resonance or to track multiple bands [Fig. 4].

The measured response curves can be analyzed in a variety of ways. The most straightforward is to individually fit each to the simple harmonic oscillator model [Eq. (1)] to determine the resonant frequency, amplitude, and $Q$-factor at each point and display each as 2D images and/or use as a feedback signals. This fast Fourier transform/fitting routine substitutes the traditional lock-in/low-pass filter that provides amplitude and phase at a single frequency. In the BE method, parallel acquisition of the response at all frequencies within the band allows complete spectral acquisition at ~10 msec/pixel rate, well within the limit for SPM imaging. Thus, in the BE response the system is excited and the response is measured simultaneously at all frequencies within the excited band (parallel detection), maximizing the signal/noise ratio.

This feature of BE is most obvious in comparison with the lock-in detection. For lock-in homodyne detection, the optimal sampling of the system response can be achieved only if the residence time at each frequency point is $\delta\tau > Q/\omega$ (Fig. 3). Therefore, sampling of the full amplitude-frequency curve requires a time of $NQ/\omega$, where $N$ is number of frequency points. For $N = 100$, $Q = 500$ (typical for ambient non-contact modes), and $\omega = 2\pi\,100$ kHz, this yields a minimum time for a lock-in of ~80 ms/pixel. Most lock-in amplifiers have additional time constants associated with input and output filters, which can add 0.5 – 5 ms/frequency point, equivalent to ~ 100 ms/pixel. This translates to acquisition times of ~ 4 hours for standard 256x256 image. Hence, compared to standard lock-in detection, the BE approach allows a time reduction for acquiring a sweep by a factor of 10 to 100 per pixel by avoiding the requirement for the $Q/\omega$ delay (or, rather, by performing this detection on all frequencies in parallel). Notably, the BE acquisition time does not depend on the width of the frequency band, or, equivalently, the number of frequency points (unlike lock-in detection, which scales linearly), allowing both for large "survey" scans in frequency space to detect relevant features of a system response (primary resonances), and precise probing of the behavior in the vicinity of a single resonance.

### IV. Implementation of BE SPM

In the BE method, the cantilever is tuned using a standard SPM fast tuning procedure to determine the corresponding resonant frequency. The frequency band is chosen such that the resonance corresponds approximately to the center of the band. In this work, we used a signal having uniform amplitude within the band, even though more complex frequency spectra can be used, as shown in Fig. 4. A typical example of an excitation and response signal in Fourier and time domains in standard SPM and BE SPM are shown in Fig. 5.

The BE signal is synthesized prior to image acquisition and then downloaded to an arbitrary waveform generator and used to drive the tip either electrically (such as in Piezoresponse Force Microscopy and KPFM), mechanically (as in tapping mode atomic force microscopy, Magnetic Force Microscopy, and Electrostatic Force Microscopy), or to drive an external oscillator below the sample (Atomic Force Acoustic Microscopy). The response signal is acquired using a fast data acquisition card (NI-6115) and Fourier transformed to yield amplitude-frequency and phase-frequency curves. The ratio of the Fourier transforms of the response and excitation signal yield the transfer function of the system within the selected band. The amplitude-frequency and phase-frequency curves in each point are stored as 3D data arrays for subsequent analysis.

The data at each pixel is fitted to the simple harmonic oscillator (SHO) model Eq. (1). The fitting yields the local response, $A_i^{\max}$, resonant frequency $\omega_{i0}$, and $Q$-factor (or dissipation). The fitting can be performed either on amplitude or phase data, or simultaneously on both. To ensure adequate weighting, in the latter case the data is transformed into real and imaginary parts, $A\cos\varphi$ and $A\sin\varphi$. The derived SHO coefficients are plotted as 2D maps. Note that more complex forms of data analysis [using different physical models,[21] statistical fits, wavelet signal transforms,[22] etc.] are possible.

The BE method for a single point can then be extended to spectroscopy and imaging in the point-by-point, and line-by-line modes. In spectroscopic BE measurements, the waveforms are applied to the probe and the response is measured as a function of a slowly varying external parameter (tip-surface separation, force, or bias) at a single point of the surface to yield 2D spectroscopic response-frequency-parameter maps (spectrograms). Subsequent fitting using SHO model allows 1D response-parameter spectra (e.g., dissipation-distance or response-distance curves) to be extracted and compared with the varying parameter (such as force-distance data).

In point-by-point measurements, the tip approaches the surface vertically until the deflection set-point is achieved. The amplitude-frequency data is then acquired at each point. After acquisition, the tip is moved to the next location. This is continued until a mesh of evenly spaced $M$ x $N$ points is scanned to yields 3D data array. Subsequent analysis yields 2D maps of corresponding quantities.

In line-by-line measurements, the BE signal substitutes the standard driving signal during the interleave line on a commercial SPM (MultiMode NS-IIIA). The topographic information in the main line is collected using standard intermittent contact or contact mode detection. The data is processed using an external data acquisition system and is synchronized with the SPM topographic image to yield BE maps.

## V. Force-Distance and Voltage Spectroscopy with BE-SPM

In the following, we illustrate the applicability of the BE method to several specific SPM applications including (i) point force- spectroscopy, (ii) bias- spectroscopy, and (iii) imaging of magnetic dissipation. As an illustration of point force spectroscopy, BE-mapping of the frequency dependence of the cantilever response with tip-surface separation under an electrostatic driving force is illustrated in Fig. 6 (a). The measurements are performed on freshly cleaved mica surface in ambient using gold-coated tips (Micromasch, $k$ = 1 N/m). On approaching the surface (bottom to top) the response gradually increases due to an increase of capacitive forces, while the resonance frequency remains constant (Region I). In close vicinity to the surface, the resonant frequency decreases due to strong attractive interactions (inset). A rapid change in the resonant structure occurs upon transition from the free to bound cantilever modes (jump to contact). Upon increasing the contact force by loading the cantilever, a slight increase in contact stiffness is observed (Region II). The reverse sequence is observed during retraction (Region III). The total acquisition time for this data is 100 s. The individual resonances at points along the vertical tip trajectory can be fitted by the SHO model and the resulting evolution of amplitude, resonant frequency, and dissipation are shown in Fig. 6 (b). These data illustrate BE spectroscopy of the dissipation in the near-surface layer and bulk material.[23] In these cases, the increased damping for small interaction forces is due to the relatively larger contribution of the surface layer to the overall contact. Note that BE allows the probing of extremely broad frequency range (25 – 250 kHz) in ~ 1 s – a comparable scan using lock-in at single frequency would require ~30 min.

A second example of the BE method is the voltage spectroscopy of dynamic processes in ferroelectric materials. Here, the dynamic response of the electrically driven, conductive cantilever in contact with a ferroelectric LiNbO$_3$ surface is measured as a function of dc bias on the tip, as illustrated in Fig. 7 (a), with a total acquisition time of 100 s. The response amplitude, resonance frequency, and quality factor are shown in Fig. 7 (b). The rapid decrease in amplitude and quality factor outside the -50 V < $V_{dc}$ < 50 V interval is associated with the nucleation of ferroelectric domains and the opening of additional damping channels due to the motion of the newly formed ferroelectric domain walls. The formation of a domain can be observed in the standard PFM image after data acquisition [inset in Fig. 7 (a)] (note that experiment was performed twice, giving rise to two domains).

## VI. Imaging Magnetic Dissipation with BE-MFM

The BE method is universally applicable for SPM provided that sufficient sampling of the resonance curve can be achieved for point spacing in frequency domain $\Delta f = 1/T$, where *T* is the pulse duration (time per pixel). At the same time, an arbitrarily broad frequency band can be excited, at the expense of the signal to noise ratio. Practically, these considerations favor the BE method for the systems with high resonance frequencies and moderate Q-factors (10-100), corresponding to the resonant peak width > ~0.3 kHz. Experimentally, most contact mode techniques have high resonant frequency (the first contact mode resonance is ~ 4 times the free resonance) and lower *Q*-factors. Similarly, imaging in liquid is typically associated with low *Q*-factors (~1-20, as compared to 100-600 in air). Hence, to demonstrate the universal applicability of the BE method for ambient and liquid SPMs, we have chosen

Magnetic Force Microscopy (MFM) as a prototype model with relatively low resonance frequency (~50-100 kHz) and high quality factors (typically ~200).

BE-MFM was implemented in a standard line-by-line interleave mode with intermittent contact mode feedback for topographic detection. As a model system for magnetic dissipation measurements, we have chosen Yttrium- Iron garnet (YIG, $Y_3Fe_5O_{12}$), dissipation in which have been studied previously by conventional Magnetic Dissipation Force Microscopy,[11] i.e. phase detection in MFM using Cleveland method for analysis of phase data. A large scale MFM image is shown in Fig. 8. The images exhibit a "flower-like" magnetic domain pattern characteristic for this material. Superimposed on the pattern are small circular features corresponding to dissipation at the defect centers in YIG.[24] The MDFM images obtained with standard sinusoidal driving[11] in all cases show the superposition of domain and dissipative contrasts. As discussed above, this is a consequence of the detection mechanism in which changes in Q-factor cannot be probed reliably without calibration of the probe transfer function.[18]

In BE-MFM, the standard, sinusoidal excitation and phase-locked loop frequency detection is substituted for direct acquisition of the response in the predefined frequency interval. Shown in Fig. 9 (a-d) are the surface topography, amplitude, quality factor, and resonance frequency BE MFM images of the YIG surface. Note that the amplitude image shows only weak variation across the surface, as expected. The frequency shift image shows flower-like patterns with high contrast, similar to standard MFM with frequency tracking. The *Q*-factor image illustrates the characteristic circular features due to magnetic dissipation.

The dissipation power at the defect compared to the sample surface can be estimated as $P_{tip} = kA^2\omega_0 \Delta Q / 2Q^2 \approx 0.012$ pW. In this estimate, the effective amplitude is scaled by the intensity factor, $I_{\text{det}}$, taking into account the response decay away from the resonance. Experimental noise in the $Q$-factor image is 0.8. The theoretical thermal limit is $\delta Q^2 = 2k_B T Q^3 B / k A^2 \omega_0$, where $k_B$ is Boltzmann's constant and $B$ is the measurement bandwidth.[10] For our case ($k$ = 2.55 N/m, $A$ = 9 nm, $\omega_0$ = 75 kHz, $Q$ = 199) the thermomechanical limit on dissipation sensitivity is $\delta Q$ = 0.14 for a bandwidth of 33 Hz. Hence, the BE method allows dissipation detection with sensitivity close to the thermomechanical noise of the cantilever.

Note that unlike conventional MFM, the amplitude, resonant frequency, and dissipation are measured independently, thus achieving independent determination of the three SHO parameters. Furthermore, this approach can be extended to most ambient and liquid SPM techniques, including electrical imaging by Kelvin probe force microscopy (KPFM) and electrostatic force microscopy (EFM), acoustic imaging by AFAM, and electromechanical imaging by PFM.

## VII. Summary

To summarize, we have developed an approach for dissipation imaging and transfer function determination in SPM based on a digitally synthesized excitation signal having a finite amplitude density in a predefined frequency range. This approach allows direct probing of the $Q$-factor of the cantilever, avoiding the limitations of standard lock-in detection. The applicability of the BE approach is demonstrated for mapping energy dissipation in MFM, mechanical and electromechanical probes, including loss processes during ferroelectric domain formation, and the evolution of dynamic behavior of the probes during force-distance curve acquisition. These examples illustrate the universality of the BE method, which can be used as an excitation and control method in all ambient and liquid SPM methods, including standard intermittent mode topographic imaging, magnetic imaging by MFM, electrical imaging by KPFM and EFM, acoustic imaging by AFAM, and electromechanical imaging by PFM. In these techniques, BE allows direct measurement of previously unavailable information of energy dissipation in magnetic, electrical, and electromechanical processes.

The capability of mapping local energy dissipation on the nanoscale is an enabling technology that will open a pathway towards atomistic mechanism of dissipation and establish relationships between dissipation and structure. This, in turn, will eventually allow the development of strategies to minimize and avoid undesirable energy losses in technologies as diverse as electronics, information technology, and energy storage, transport, and generation.

Furthermore, energy dissipation measurements will open a window into energy transformation mechanisms during fundamental physical and chemical processes. Simple estimates suggest that at room temperature the estimated detection limit in BE method as limited by thermomechanical noise is $P_{tip}^2 = k_B T k A^2 \omega_0 B / Q$, corresponding to ~ 0.5 fW, or ~31 mV/oscillation level for ambient environment (as compared to currently demonstrated 20 fW).[11] At low temperatures or in a high-$Q$ environment, detection of single optical phonon generation in the tip-surface junction is possible, providing information on the dissipative processes with broad applicability to nanomechanics and nanotribology. The implementation of BE method at low temperatures holds the promise of an even further increase of sensitivity to the level that a single quasiparticle can be detected, providing insight into the fundamental physics of strongly correlated oxide materials and other systems on the forefront of research.

## VIII. Acknowledgements

Research sponsored by ORNL SEED funding (SJ and SVK) and in part (BJR, APB) by the Division of Materials Sciences and Engineering and Office of Basic Energy Sciences, U.S. Department of Energy, under contract DE-AC05-00OR22725 at Oak Ridge National Laboratory, managed and operated by UT-Battelle, LLC. The band excitation method is available as a part of the user program at the Center for Nanophase Materials Science, Oak Ridge National Laboratory (contact – SVK).

**Figure Captions**

**Fig. 1.** Single frequency measurements (red dots) are not always adequate for determining Q. (a) For a constant driving force, the amplitude (peak height) is inversely proportional to the quality factor (peak width) of the system. In such a case, dissipation can be determined from amplitude at a single frequency (e.g. at the resonance). (b) For a non-constant driving force however, the amplitude and dissipation are independent. Hence, probing energy dissipation requires measuring the response over a range of frequencies across a resonance.

**Fig. 2.** Operational principle of the BE method in SPM. The excitation signal is digitally synthesized to have a predefined amplitude and phase in the given frequency window. The cantilever response is detected and Fourier transformed at each pixel in an image. The ratio of the FFTs of response and excitation signals yield the cantilever response (transfer function). Fitting the response to the simple harmonic oscillator yields amplitude, resonance frequency, and *Q*-factor that are plotted to yield 2D images, or used as feedback signals.

**Fig. 3.** Lock-in sweep detection. (a) One type of excitation signal (chirp) can be represented as (b) a sinusoidal excitation with linearly varying frequency. (c) Standard lock-in probes operate at a single frequency and can be represented as a band pass filter of width $\Delta f \sim 1/\tau$. Hence, the lock-in sweep works as a moving bandpass filter. BE detects the response at all frequencies simultaneously. (d) Envelope of the response at a given frequency. The linearity implies that the system responds at the same frequency as the excitation signal. However, due to the finite quality factor, response is not instant and response amplitude increases from 0 to the saturated value in a time on the order of $Q/\omega_0$. Similarly, response persists after initial excitation. Even in the ideal case (perfect notch filter) lock-in detection loses information in the shaded region (response after excitation is turned off). The BE method utilizes the full frequency domain of the excitation, avoiding this dynamic effect. Note that the two methods are equivalent for low-Q systems.

**Fig. 4.** Frequency spectrum of excitation signal in single frequency (a) static and (b) frequency-tracking cases. In frequency tracking methods, the excitation frequency and excitation amplitude are varied from point to point. In the band excitation method, the response in the selected frequency window around the resonance is excited. The excitation signal can have (c) uniform spectral density [as is the case in this paper], or (d) increased spectral density on the tails of resonance peak to achieve better sampling away from the peak. (e) The resonance can be excited simultaneously over several resonance windows. Also, (f) the phase content of the signal can be controlled, for example, to achieve *Q*-control amplification. The excitation signal can be selected prior to imaging, or adapted at each point so that the center of the excitation window follows the resonance frequency (c) or phase content is updated (f). This is important for e.g. contact mode techniques, when the tip-surface contact area and hence the resonance frequency changes significantly with position.

**Fig. 5.** Excitation (blue) and response (red) signals in standard SPM techniques in (a) time domain and (b) Fourier domain. Excitation (blue) and response (red) signals in BE SPM in (c) time domain and (d) Fourier domain. In BE, the system response is probed in the specified

frequency range (e.g., encompassing a resonance), as opposed to a single frequency in conventional SPMs.

**Fig. 6.** (a) Evolution of the dynamic properties of the cantilever-surface system during force-distance curve acquisition. (b) Deconvolution of the BE data in amplitude, resonant frequency, and Q-factor measured along the force-distance curve.

**Fig. 7.** (a) Bias dependence of the amplitude-frequency response curve for the tip in contact with a ferroelectric $LiNbO_3$ surface. The inset shows the domain formed by the end of the BE measurement. (b) Bias dependence of amplitude, resonant frequency, and Q-factor. The saturation of electromechanical response and decrease in Q-factor evidence the onset of ferroelectric switching, which opens an additional channel for energy dissipation.

**Fig. 8.** Standard magnetic force microscopy (a) amplitude and (b) phase images of the YIG surface. The amplitude image shows clear "flower-like" pattern related to the presence of magnetic domains. Phase image shows rings due to the dissipation energy losses at the magnetic dissipation centers. Note that both images illustrate both domain-related and dissipation-related features due to cross-talk.

**Fig. 9.** (a) Surface topography, (b) response amplitude, (c) resonance frequency, and (d) Q-factor image of YIG surface obtained in BE MFM. The ring-like structures form due to magnetic dissipation centers as corroborated by conventional MDFM. The frequency and Q-factor images illustrate complete decoupling between the force gradient (frequency shift) and dissipation (*Q*-factor) data. (c) Average amplitude curve, local amplitude curve and difference between the two at point 1, note the asymmetry. (d) Average amplitude curve, local amplitude curve and difference between the two at point 2, note the drop in amplitude. Vertical scale for (a) is 2 nm.

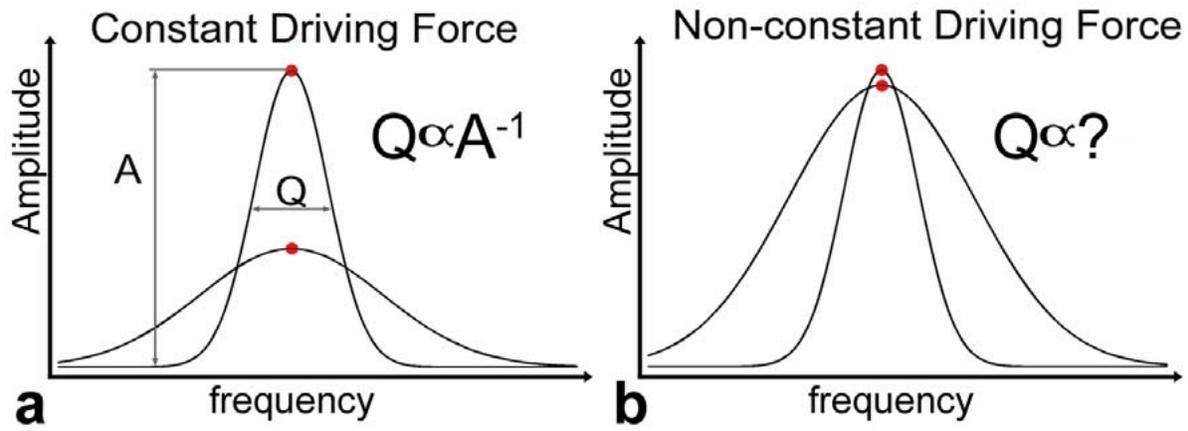

**Figure 1.** Stephen Jesse, Sergei V. Kalinin *et al.*

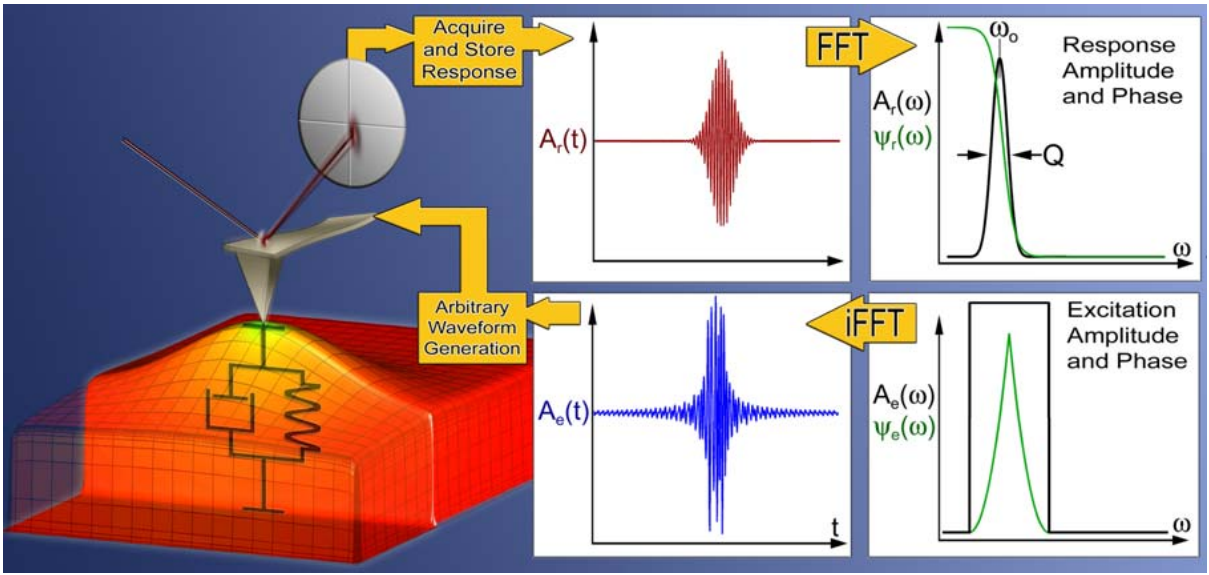

**Figure 2.** Stephen Jesse, Sergei V. Kalinin *et al.*

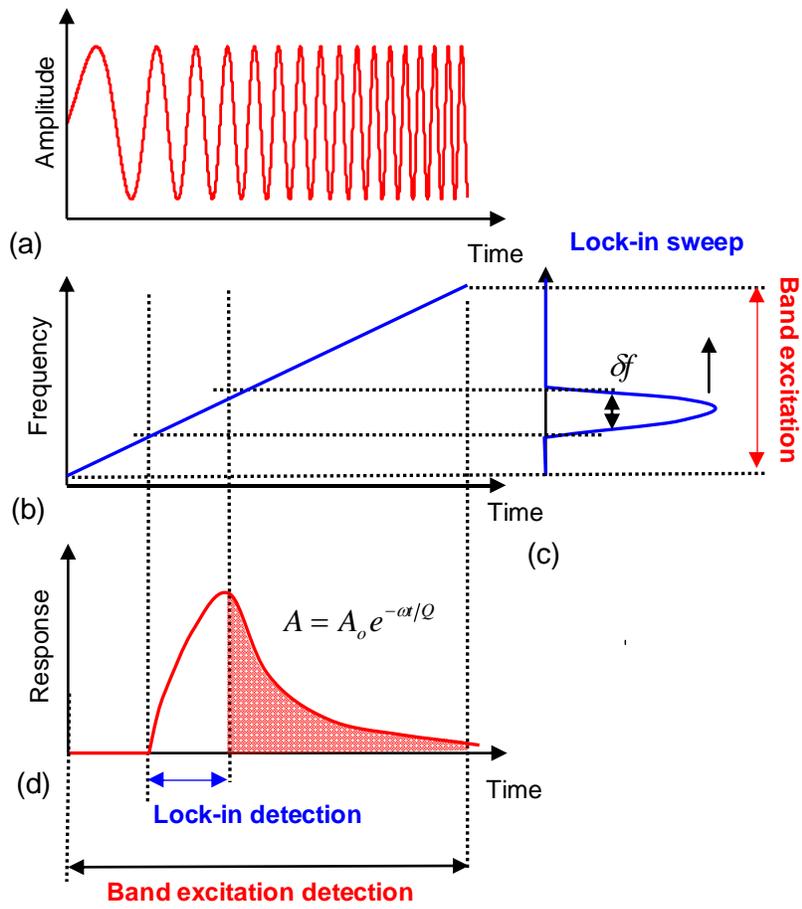

**Figure 3.** Stephen Jesse, Sergei V. Kalinin *et al.*

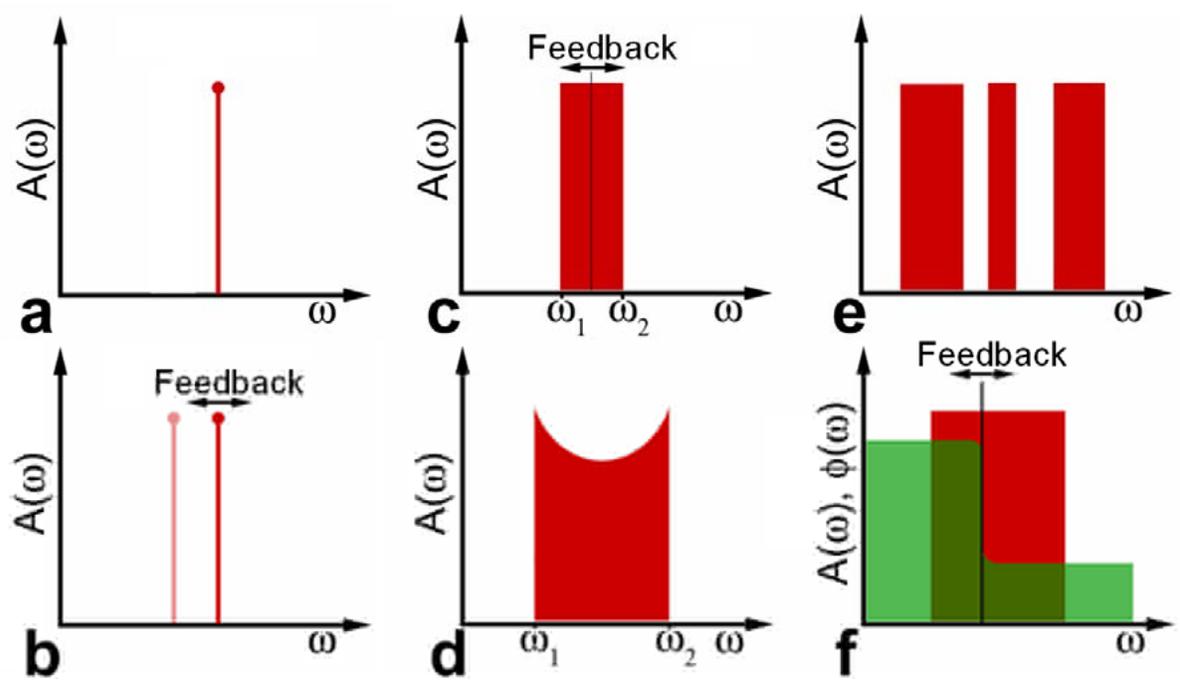

**Figure 4.** Stephen Jesse, Sergei V. Kalinin *et al.*

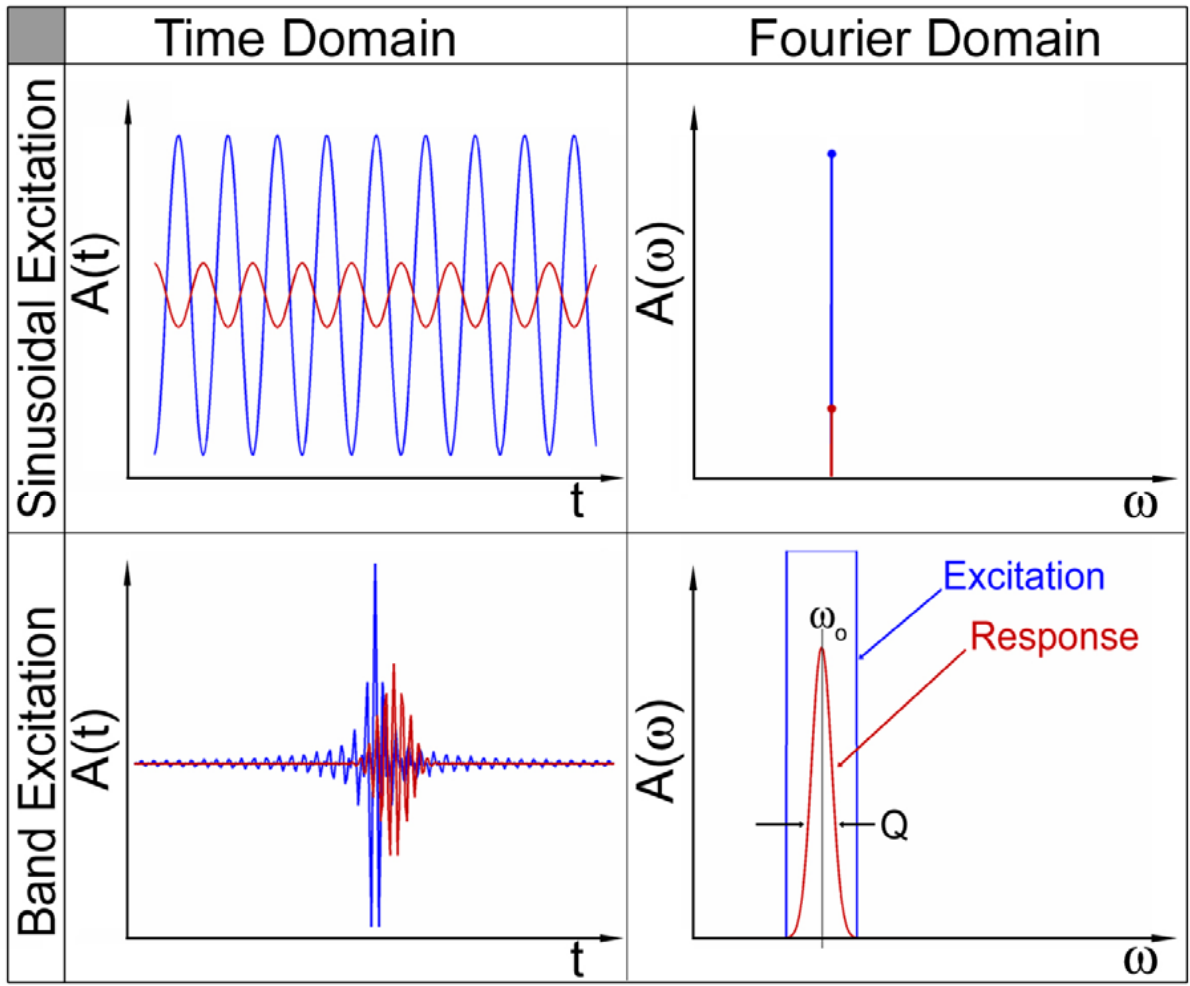

**Figure 5.** Stephen Jesse, Sergei V. Kalinin *et al.*

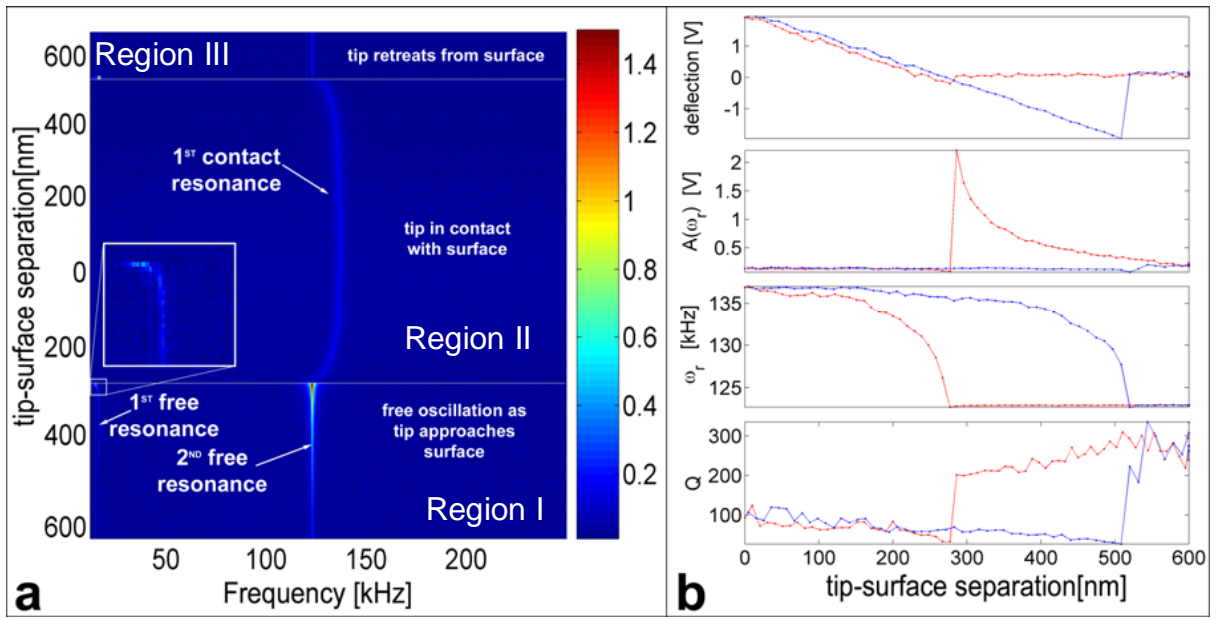

**Figure 6.** Stephen Jesse, Sergei V. Kalinin *et al.*

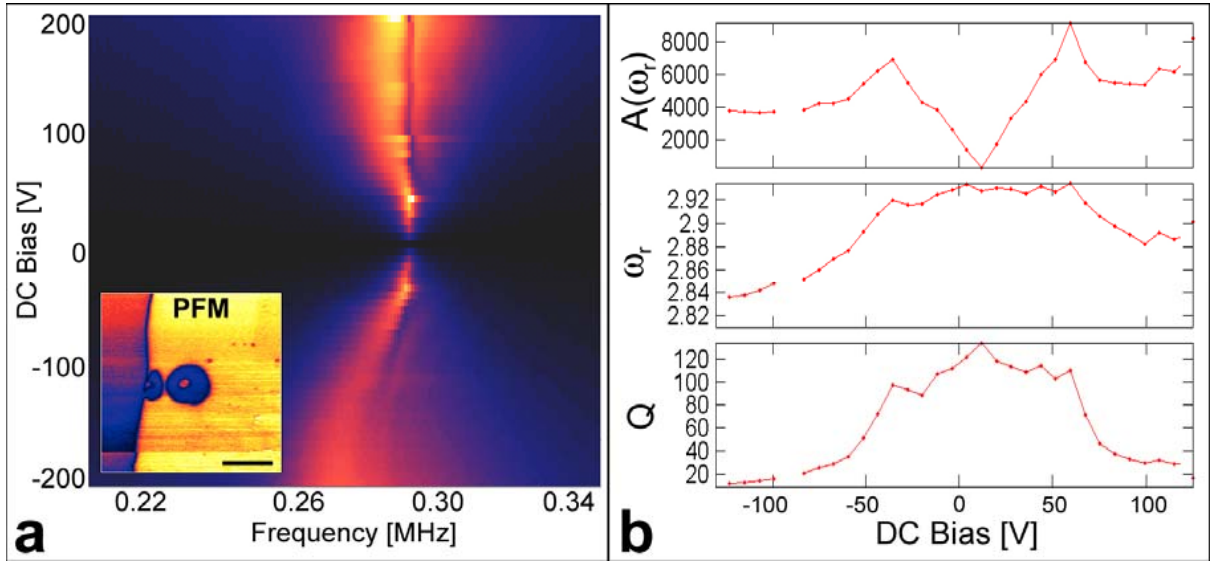

**Figure 7.** Stephen Jesse, Sergei V. Kalinin *et al.*

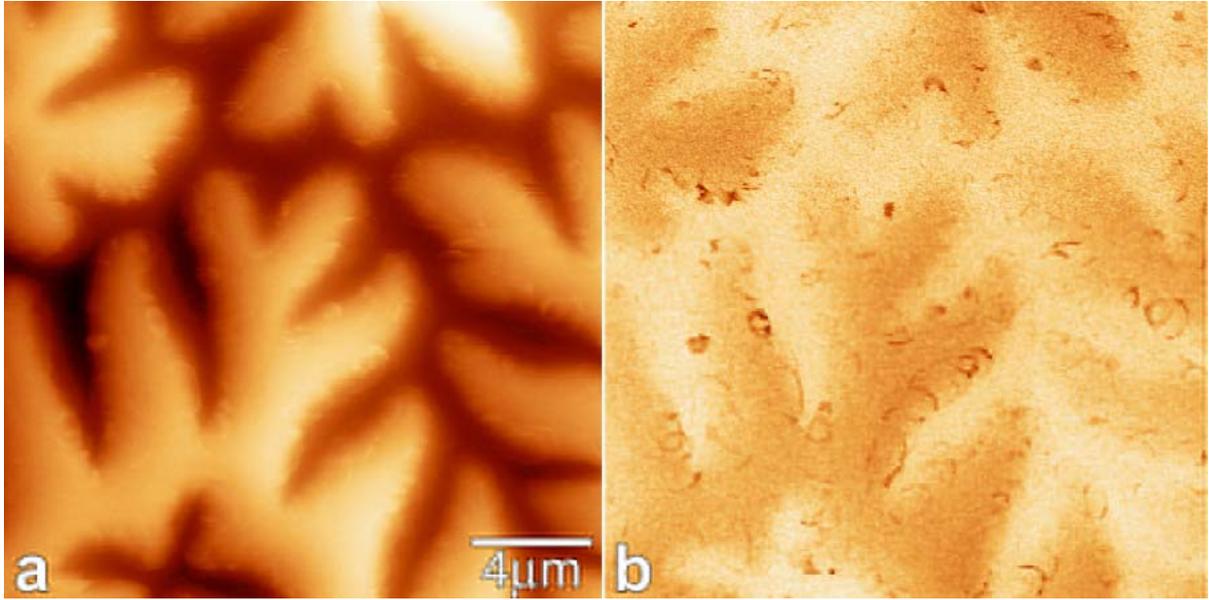

**Figure 8.** Stephen Jesse, Sergei V. Kalinin *et al.*

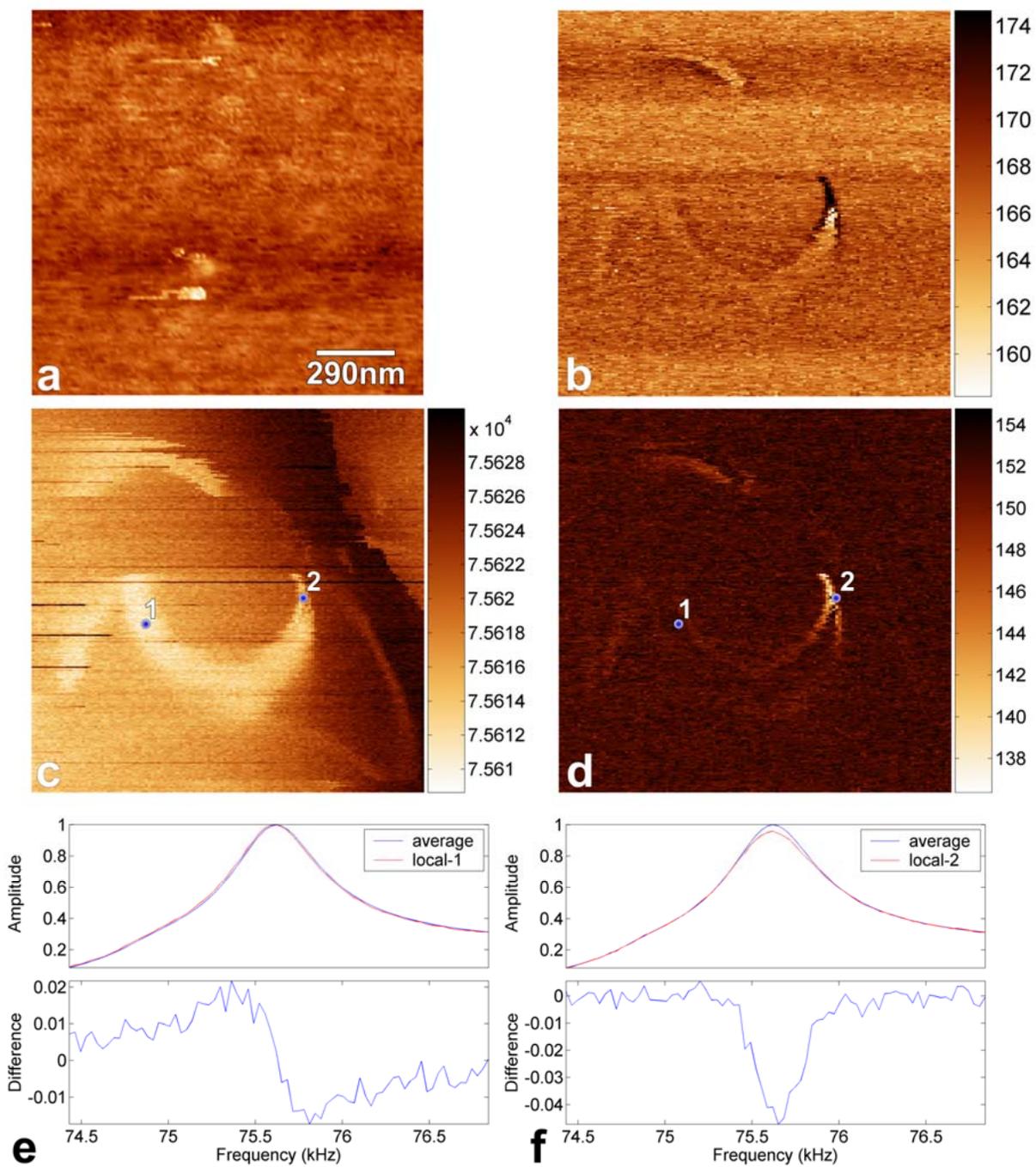

**Figure 9.** Stephen Jesse, Sergei V. Kalinin *et al*.